\documentclass[AMA,STIX1COL]{WileyNJD-v2}

\articletype{Article Type}%

\usepackage{subfigure}
\usepackage{url}
\usepackage{colortbl}
\usepackage{xcolor}
\usepackage{multirow}
\usepackage{multicol}
\usepackage{xcolor,colortbl}
\usepackage{endfloat}

\begin{document}
\newcommand{\mc}[2]{\multicolumn{#1}{c}{#2}}
\definecolor{Gray}{gray}{0.85}
\definecolor{LightCyan}{rgb}{0.88,1,1}

\newcolumntype{a}{>{\columncolor{Gray}}c}
\newcolumntype{b}{>{\columncolor{white}}c}

\title{Rank Selection for Non-negative Matrix Factorization}

\author[1]{Yun Cai*}

\author[2]{Hong Gu}

\author[3]{Toby Kenney}

\authormark{Yun Cai \textsc{et al}}

\address[1]{\orgdiv{Department of Mathematics and Statistics}, \orgname{Dalhousie}, \orgaddress{\state{Halifax}, \country{Canada}}}

\address[2]{\orgdiv{Department of Mathematics and Statistics}, \orgname{Dalhousie}, \orgaddress{\state{Halifax}, \country{Canada}}}

\address[3]{\orgdiv{Department of Mathematics and Statistics}, \orgname{Dalhousie}, \orgaddress{\state{Halifax}, \country{Canada}}}

\corres{*Yun Cai, Department of Mathematics and Statistics, Dalhousie. \email{yn298893@dal.ca}}


\abstract[Summary]{Non-Negative Matrix Factorization (NMF) is a widely used dimension reduction method that factorizes a non-negative data matrix into two lower dimensional non-negative matrices: One is the basis or feature matrix which consists of the variables and the other is the coefficients matrix which is the projections of data points to the new basis. The features can be interpreted as sub-structures of the data. The number of sub-structures in the feature matrix is also called the rank which is the only tuning parameter in NMF. An appropriate rank will extract the key latent features while minimizing the noise from the original data. In this paper, we develop a novel rank selection method based on hypothesis testing, using a deconvolved bootstrap distribution to assess the significance level accurately despite the large amount of optimization error. In the simulation section, we compare our method with a rank selection method based on hypothesis testing using bootstrap distribution without deconvolution, and with a cross-validated imputation method \cite{lin2020optimization}. Through simulations, we demonstrate that our method is not only accurate at estimating the true ranks for NMF especially when the features are hard to distinguish but also efficient at computation. When applied to real microbiome data (e.g. OTU data and functional metagenomic data), our method also shows the ability to extract interpretable subcommunities in the data.}



\maketitle

\footnotetext{\textbf{Abbreviations:} NMF, non-negative matrix factorization; OTU, Operational Taxonomic Unit}

\section{Introduction}\label{sec1}
\subsection{Non-Negative Matrix Factorization}\label{subsec11}
Consider a non-negative data matrix $X$ with $p$ variables (rows) and $n$ observations (columns). NMF \cite{lee1999learning} attempts to approximate the matrix $X$ by the product $TW$ where $T$ is a $p\times k$ non-negative matrix and $W$ is a $k\times n$ non-negative matrix.  When $k\ll np/(n+p)$, the dimension is significantly reduced from the original problem. The matrices $T$ and $W$ are called the feature matrix and the weight matrix respectively. Each column of W represents the corresponding observation from $X$ as a non-negative linear combination of the columns of $T$.

Because of the non-negativity constraints, the features and weights selected by NMF tend to be sparse. Furthermore, the features, or types in the matrix T represent parts of the data X. This is different from other dimension reduction methods like principal component analysis or vector quantization, and makes interpretation of results easier. In many contexts, the non-negative combinations can be viewed as mixtures of the types.

NMF has been widely applied in many areas including image data \cite{lee1999learning}, automatic music transcription \cite{wu2022semi}, computational biology \cite{brunet2004metagenes} \cite{wu2022semi}, and microbiome data \cite{cai2017learning}. The microbiome refers to the whole habitat, including the microorganisms, their genomes and the environmental conditions.\cite{marchesi2015vocabulary} Microbial communities, which consists of a large number of different microorganisms, have been shown to be associated with human health and global nutrient cycling.\cite{gilbert2012defining}\cite{arrigo2005marine}\cite{fujimura2010role}\cite{sekirov2010gut}
The non-negative structure of NMF can be neatly interpreted as expressing the microbial community as a combination of subcommunities, which may represent interacting groups of microbes. Jiang\cite{jiang2012functional} has applied NMF to marine microbes data to investigate the biogeography of microbial function by extracting a small number of features associated with both microbial protein profiles and sampling location. Ko\cite{ko2021identification} used NMF to estimate functional microbial interactions from microbiome data.

The approximation $X\sim TW$ can be fitted to minimize either the squared error loss, or more generally the negative log-likelihood of $X$ under some assumed parametric distribution. There are multiplicative algorithms for fitting NMF both for squared error loss and for log-likelihood for Poisson distributed $X$ \cite{seung2001algorithms}. These methods are implemented by the R package NMF, which we use for fitting NMF in this paper. While these methods converge robustly, the solution to NMF is not always unique \cite{laurberg2008theorems}, and even if it is unique, the likelihood surface can be multimodal \cite{chu2004optimality}, meaning that the optimization can converge to a local optimum. This creates difficulties that we need to overcome in our method. 

The only tuning parameter in the NMF algorithm is the rank of the feature matrix, $k$, which is the number of features. The interpretation of $k$ is the number of underlying sub-structures or subcommunities extracted from the data. The selection of $k$ not only affects the performance of NMF but also relates to the interpretation of NMF results. Too small $k$ may lead to key features missing and too large $k$ value may have overfitting issues. It is therefore necessary to select an appropriate $k$ value for the data. The purpose of this paper is to develop a new method for selecting this rank $k$.

\subsection{A brief literature review for rank selection in NMF}\label{subsec12}
 The usual approach to NMF rank selection is to use expert knowledge \cite{gillis2014and}. However, for many analyses in practice, no-one has the insight needed to select the rank. Even if there are experts with a good intuition of what rank is appropriate, being able to support the choice with statistical evidence is valuable.

 When the expert knowledge is unavailable, a popular approach for rank selection is to compute some model measures for a range of rank values and choose the rank value according to the measure's criteria. Cophenetic correlation coefficient and Dispersion are the most commonly used measurements \cite{muzzarelli2019rank}. The Cophenetic correlation coefficient method measures the reproducibility of the clustering of observations based on the weight matrix of NMF for many runs with different initial values. Because of the multimodality of the likelihood or squared errors, runs of NMF with different initial values can produce different results. The idea is that for the appropriate number of features, there should be a clear optimum, so runs with different initial values should give the same solution. If the number of features is too large, then the optimal likelihood or mean squared error will vary between runs, leading to a smaller Cophenetic correlation. The method selects the rank value where the Cophenetic correlation has a steep drop off \cite{brunet2004metagenes}. The Dispersion method uses a similar procedure to the Cophenetic coefficient method but calculates Dispersion coefficients instead \cite{kim2007sparse}. These two methods both measure the robustness of the clustering analysis results, instead of the goodness-of-fit of the NMF results to the data, thus could lead to suboptimal solutions \cite{frigyesi2008non}. Other ideas include comparing residuals, sparseness and Description Length \cite{squires2017rank}. These methods are all ad hoc in that they select the rank according to the plots based on intuition instead of a solid statistical inference results. The results will be inaccurate when a clear drop off point is not available.
 
 Another popular approach is to use cross-validation. For example, using 2-fold cross-validation to split the data randomly into two halves and apply NMF to the two parts separately. The rank minimizing the average reconstruction error between the two parts is selected for NMF \cite{sotiras2017patterns}.  Lin\cite{lin2020optimization} introduces a rank selection method based on imputation. They randomly delete $30\%$ of the data entry from the original data each time and treat them as missing values in the NMF procedure. The rank is chosen by minimizing the mean square errors of the reconstruction of the missing entries. Their NMF algorithm to handle missing values is available in the R package NNLM \cite{lin2020optimization}. Compared with the Cophenetic correlation coefficient method, Dispersion method and 2-fold cross validation method, the NNLM imputed mean squared error method estimates the rank more accurately when applied to Normal synthetic data with different properties \cite{muzzarelli2019rank}.

There are also some other rank selection methods such as the Bayesian method \cite{hoffman2010bayesian} and singular value decomposition. The Bayesian method requires an assumption that the rank should be small \cite{hoffman2010bayesian}, which is not necessarily true for real data. Singular value decomposition chooses the rank where the singular values become small \cite{gillis2014and}. The challenge of the method arises when there isn't a clear drop to zero singular value among all ranks. Moreover, the rank of singular values is not closely related to the rank of non-negative matrix factorization. 

In this paper, we develop a goodness-of-fit test based on a deconvolved bootstrap distribution and use the test to choose an appropriate rank for NMF. The application of deconvolution to bootstaps with optimisation error appears to be novel. We then compare our rank selection method with bootstrap distribution without deconvolution and NNLM on a range of simulated data and real data. We find that our method produces accurate and stable estimation of ranks for NMF.

\section{Methods}\label{sec2}

Our rank selection for NMF is based on sequentially performing the following hypothesis test:

$H_0$: the rank of the feature matrix is $k$.

$H_a$: the rank of the feature matrix is at least $k+1$.

After applying the goodness-of-fit test, if $H_0$ is rejected, let $k=k+1$ and repeat the test until $H_0$ can not be rejected at the given significance level.

\subsection{Likelihood ratio test}\label{subsec21}

Suppose $L(k)$ is the Likelihood of the factorization using rank $k$ NMF and $L(k+1)$ is the Likelihood of the rank $k+1$ NMF results. We construct a likelihood ratio test. Under null hypothesis $H_0$, the test statistic is given by:
\begin{eqnarray}
\lambda_{LR}&=& -2\ln \left(\frac{\sup L(k)}{\sup L(k+1)}\right)\nonumber\\
 &=& -2(\ln{({\sup} L(k))}-\ln{({\sup}L(k+1)))}\nonumber\\
\end{eqnarray}

Here $\sup$ denotes the supremum and $\ln$ represents natural logarithm. 

According to standard statistical theory, under certain conditions, the likelihood ratio statistics asymptotically follow a chi-squared distribution. However, because of the non-negative boundary, the conditions for the asymptotic null chi-squared distributions are not met.  Moreover, due to the computational difficulty, it is most common that the global maximization of the likelihood is not achieved by applying the NMF algorithm with a single set of initial values. Thus, the results of a given run of the NMF algorithm are subject to computational measurement errors for the likelihood ratio statistic. For the first difficulty, we can use a bootstrap method to compute a correct null distribution. However, the computational measurement error in the likelihood ratio statistic for the bootstrap samples makes the hypothesis testing method lose its power due to the long tail in the null distribution. We develop a method to get a measurement error deconvolved test to improve the power for the likelihood ratio test in NMF rank selection. This work is also an application of our previous work on a general method for measurement error deconvolution density estimation using penalized likelihood \cite{cai2017learning}.

\subsection{Direct testing with bootstrapped null distribution (Boot-test)}\label{subsec22}

For our hypothesis test, we can use a parametric bootstrap to obtain the null distribution of the test statistic. Under the null hypothesis, the data is a realization of a random sample from a distribution $F$ with mean a linear combination of the $k$ features. The $k$ features and their coefficients are estimated by rank $k$ NMF of the original data. In the bootstrap, samples of random datasets are drawn from $\hat{F}$. These samples have the same dimension as the original data. When the data is Poisson distributed, the mean is the only parameter of the distribution. For Normal data, assuming each entry of the data is independently Normally distributed and shares the same variance, both mean and variance need to be estimated from the original data. The variance can be estimated from the residuals of the $k$ rank NMF model. For example, if the rank $k$ NMF gives $X \approx T_{p\times k}W_{k\times n}$. $TW$ is the estimated mean parameter for the Poisson or Normal distributions. The variance of the Normal distribution can be estimated by $\frac{1}{np}\sum_{i,j}(X-TW)_{ij}^2$. The negative entries in the bootstrapped sample are replaced by 0 for Normal distributed data. We apply rank $k$ and rank $k+1$ NMF on each bootstrapped sample to get the test statistic $\lambda^*$, from which the empirical null distribution of $\lambda$ can be obtained. 

Unfortunately, this hypothesis test with bootstrapped null distribution does not perform well at selecting the rank of NMF since the likelihood surface for NMF is often multimodal, and it is common for the algorithm to converge to a suboptimal factorization. This creates noise in the bootstrapped null distribution, thus reducing the power of the test. 

To improve the performance, we can rerun the algorithm with multiple starting values on the original data to choose the best log-likelihood. The mean $\hat{F}$ is estimated as the result of NMF that has the best log-likelihood. For Normal data, the variance is the average of variances estimated from each run of NMF to avoid overfitting. We also run NMF with multiple initial values on each bootstrap sample and choose the best log-likelihoods to build the null distribution. The procedure is shown in Algorithm \ref{alg1}.

\begin{algorithm}
\caption{Pseudocode for Boot-test}\label{alg1}
\begin{algorithmic}
\State $k \gets 0$
\While{p-value is significant}
\State $k \gets k+1$
\For {m different initial values}
\State apply rank $k$ NMF and rank $k+1$ NMF to the original data
\EndFor
\State $l_0(k) \gets$ largest loglikelihood of all rank $k$ NMF results; 
\State $l_0(k+1)\gets$ largest loglikelihood of all rank $k+1$ NMF results;
\State compute $T_0$ and $W_0$ when rank $k$ NMF has loglikelihood $l_0(k)$
\State $\lambda \gets -2(l_0(k)-l_0(k+1))$
\State sample bootstrap datasets from distribution with mean $T_0W_0$ (and variance estimate if data are assumed to be Normal)
  \For {all bootstrap datasets}
      \For{m different initial values}
          \State apply rank $k$ NMF and rank $k+1$ NMF to the bootstrap data
      \EndFor
      \State compute $\lambda^*$ as negative two times the difference of the best loglikelihood of all rank k NMF results and the best loglikelihood of all rank k+1 NMF results
  \EndFor
  \State get p-value;
  \EndWhile
\end{algorithmic}
\end{algorithm}

The simulation results for this method with different numbers of initial values and different true rank values are shown in Section \ref{subsec31}.

\subsection{Testing with deconvolved bootstrap null distribution (Decon-boot-test)}\label{subsec23}

The major problem with the method described above is that to ensure a good bootstrap distribution, we need to fit each bootstrap sample with a large number of different initial values, which is computationally expensive. We therefore develop a new approach to more efficiently compute the bootstrap distribution. The idea is to treat the convergence error as measurement error in the log-likelihood ratio statistic. We can then use a well-developed deconvolution method to estimate the null distribution of the statistic. Deconvolution is a method to estimate a distribution from data with additive measurement error.
In our case, if we run only one set of initial values for NMF for each bootstrap sample for each fixed $k$, then we can directly compute the estimated maximum loglikelihood $l(k)=l_0(k) +e(k)$, where $l_0(k)$ is the globally maximized log-likelihood value (unobserved) and $e(k)$ is the measurement (convergence) error. Thus 
\begin{eqnarray}
\lambda^* &= &\lambda_0^*+e\\
&=&-2(l_0(k)-l_0(k+1)+e(k)-e(k+1)) \nonumber\\
&= &-2(l_0(k)-l_0(k+1))+\{-2[e(k)-e(k+1)]\}
\end{eqnarray}

where $\lambda_0^*$ is the true likelihood ratio statistic without convergence error and $e=-2(e(k)-e(k+1))$. The deconvolution method will estimate the distribution of $\lambda_0^*$ when $\lambda_0^*$ is not observable but $\lambda_0^*+e$ is available.

Most deconvolution approaches are based on the Fourier transformation. While its mathematical solution is very neat, estimation of the characteristic function from data is somewhat unstable, and division by possibly small Fourier coefficients for the error data can greatly magnify errors. So in this paper, we use the P-MLE deconvolution method \cite{pmledecon} which is based on maximizing a penalized log-likelihood of the data (here $\lambda^*$) with a smoothness penalty. P-MLE also has the advantage over many other competing methods that its implementation can estimate the measurement error distribution non-parametrically from a pure measurement error sample. This is important because the convergence error is unlikely to follow a standard parametric family of distributions. 

\begin{figure}[htbp]
\centering
\includegraphics[width=18cm,height=13cm]{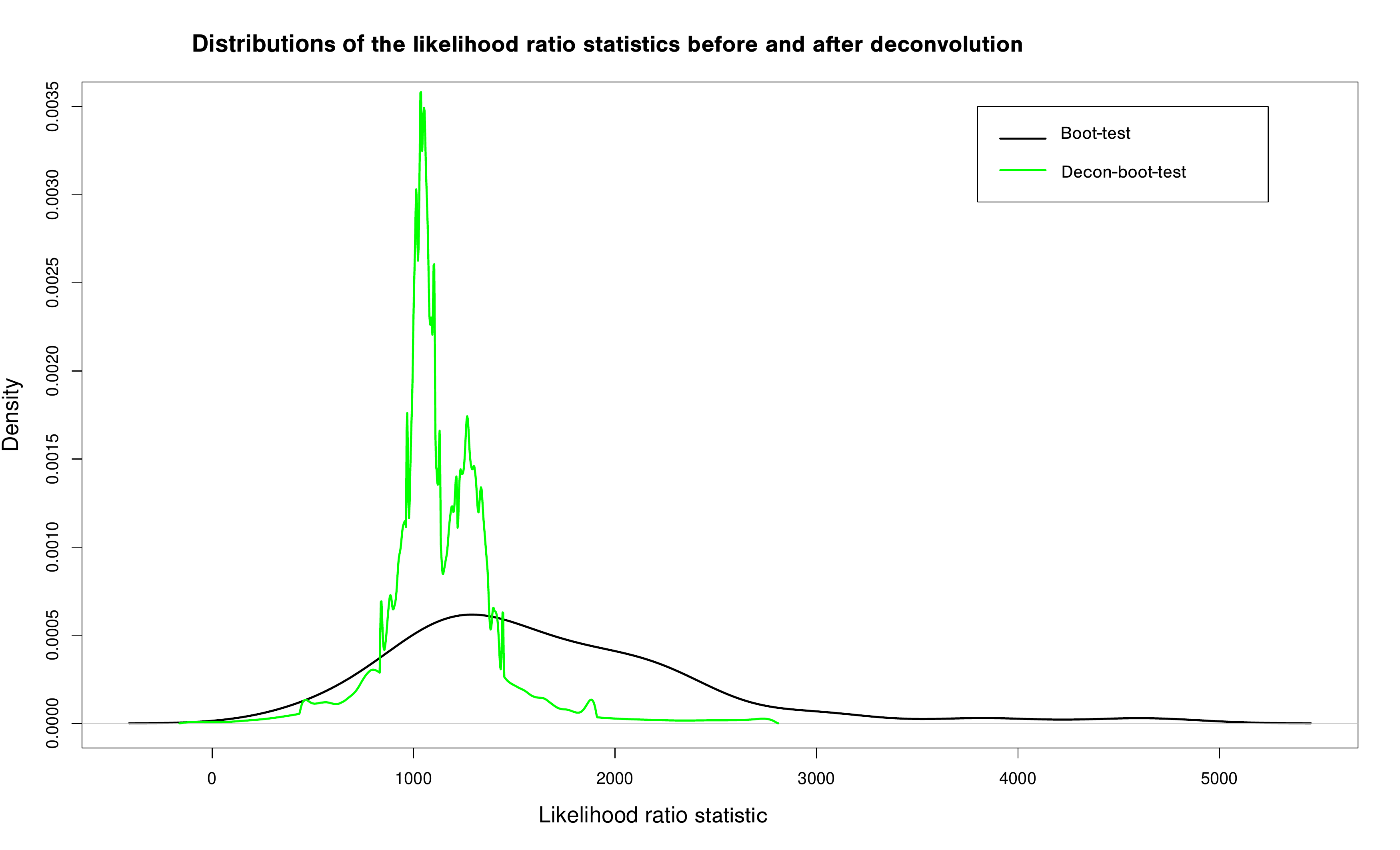}
\caption{Null distribution of the likelihood ratio statistics for rank 4 NMF vs. rank 5 NMF. The data is generated from rank 4 Poisson NMF. The black curve is the kernel estimation of bootstrapped null distribution when apply rank 4 and rank 5 NMF with single initial value for each bootstrap sample with bootstrap size as 50. The green curve is the deconvolved density of likelihood ratio statistic after applying P-MLE deconvolution method to the 50 bootstrapped Likelihood ratios.}
\label{c3F2.1}
\end{figure}

Figure \ref{c3F2.1} shows an example of bootstrapped null distribution of likelihood ratio statistic without deconvolution and after deconvolution. The deconvolved distribution of likelihood ratio statistic has a much smaller vairance than the bootstrapped null distribution of likelihood ratio statistic when NMF with single initial value applied to each bootstrap sample as convergence error is removed by deconvolution.

The sampling procedure for bootstrapped datasets is the same as in Section 2.2. However, for each sampled data, rank $k$ NMF and rank $k+1$ NMF are run only once to get $\lambda^*_0+e$. In order to get a pure error sample, we perform rank $k$ and rank $k+1$ NMF with multiple initial values for an additional bootstrap sample. Since we are trying to maximize the log-likelihood, we will assume that the largest likelihood from multiple runs of different initial values is the true maximum likelihood. If the number of different initial values is large enough, this should be close to be true. The measurement error samples of $e(k)$ and $e(k+1)$ can be obtained from the largest loglikelihood among different initial values minus the loglikelihoods of all NMF runs. The pure error sample for $e$ can be obtained as negative two times the difference between $e(k)$ and $e(k+1)$ for all different pairs of $e(k)$ and $e(k+1)$. Our method is implemented in the R package for Decon-boot-test, which is available on CRAN with the name Deconboottest at \url{https://CRAN.R-project.org/package=DBNMFrank}.

A complete description of the procedure is given in Algorithm \ref{alg2}:

\begin{algorithm}
\caption{Pseudocode for Decon-boot-test}\label{alg2}
\begin{algorithmic}
\State $k \gets 0$
\While{p-value is significant}
\State $k \gets k+1$
\For {m different initial values}
\State Applying rank $k$ NMF and rank $k+1$ NMF to original data
\EndFor
\State $l_0(k) \gets$ largest loglikelihood of all rank $k$ NMF results; 
\State $l_0(k+1)\gets$ largest loglikelihood of all rank $k+1$ NMF results;
\State compute $T_0$ and $W_0$ when rank $k$ NMF has loglikelihood $l_0(k)$
\State $\lambda \gets -2(l_0(k)-l_0(k+1))$
\State sample bootstrap datasets from distribution with mean $T_0W_0$ (and variance estimate if data are assumed to be Normal)
  \For {all bootstrap datasets}
  \State compute $\lambda^*$ by applying rank $k$ NMF and rank $k+1$ NMF on bootstrap datasets.
  \EndFor
   \For {m different initial values}
   \State compute $l_i(k)$ by applying rank $k$ NMF on a bootstrap sample;
   \State compute $l_j(k+1)$ by applying rank $k+1$ NMF on the same bootstrap sample;
   \EndFor
   \For{i $\gets$ 1 \textbf{to} m }
   \State $e_i(k) \gets (max_{\text{i}}(l_i(k)))-l_i(k)$ ;
   \EndFor
   \For{j $\gets$ 1 \textbf{to} m}
   \State $e_j(k+1)\gets (max_{\text{j}}(l_j(k+1)))-l_j(k+1)$ ;
   \EndFor
   \State $e \gets -2(e_i(k)-e_j(k+1))$ for all $i$ and $j$;
   \State compute null distribution of $\lambda^*_0$ by applying P-MLE on $\lambda^*$ and $e$;
  \State get p-value;
   \EndWhile
\end{algorithmic}
\end{algorithm}

\section{Simulation}\label{sec3}

In this section, we compare the performance of Boot-test and Decon-boot-test introduced in Sections \ref{subsec22} and \ref{subsec23} and NNLM\cite{lin2020optimization} which is based on missing data imputation, on Poisson NMF data, Normal NMF data and non-NMF structure data. NNLM minimizes Kullback-Leibler divergence loss to estimate Poisson data's factorization and square error loss to estimate Normal data's factorization. In the simulation, following the suggestions from \cite{lin2020optimization}, we randomly set $30\%$ of the data entries as NA and apply NNLM with a sequence of rank values from 1 to 50 to impute the missing data. The procedure is repeated 10 times and the rank value that minimizes the average loss (KL divergence for Poisson data and MSE for Normal data) of the imputed part is selected as the rank for the data. 

For the Boot-test and Decon-boot-test methods, we find that the bootstrap size doesn't affect the rank estimates too much on the simulation data. We only show the simulation results when bootstrap size is 50 in this section. Also, NMF with 50 different initial values consistently converges to the optimal likelihood when running from simulated data. So for the Boot-test, we follow procedure \ref{alg1} and set $m=1$, $m=3$, $m=10$, $m=30$ and $m=50$ for each dataset and each bootstrap sample. For the Decon-boot-test, we follow procedure \ref{alg2} and set $m=50$. 

We generate 50 replicates for each scenario. For the generated NMF data, the feature matrix is fixed in each scenario and the weight matrix is randomly generated in each replicate. All three methods are applied to the same data in each replicate. The significance level used for Boot-test and Decon-boot-test in the simulations is $0.1$.

\subsection{Poisson data simulation}\label{subsec31}

We simulate a range of Poisson NMF scenarios with true rank equal to 2, 4, 6, 8, 10 and 30, and a scenario where there is no NMF structure. For true rank 2, 4, 6, 8 and 10, we simulate 100 observations, while for the true rank 30, we simulate 300 observations. For the no NMF structure scenarios, we simulate 131 observations. For rank 2 and rank 4, we simulate five scenarios to assess the sensitivity of the methods when one of the types is close to a non-negative linear combination of the others. For the higher true ranks, we simulate one scenario for each true rank. 
For ranks 6, 8 and 10, all features used in the simulation are estimated from the healthy individuals in the Qin dataset\cite{qin2010human}, using Poisson NMF with the given rank. One feature for the true rank 2 simulation and three features for the true rank 4 simulation are also estimated from the healthy individuals in the Qin dataset using NMF with ranks 1 and 3 respectively. The Qin data is a human gut metagenomic dataset extracted from 99 healthy people and 25 IBD patients. 
For rank 30, the features used in the simulation are calculated by applying rank 30 NMF to Person 2's gut data in the moving picture data \cite{Caporaso}. The moving picture data is the most detailed investigation of temporal microbiome variation to date. It consists of a long-term sampling series from two human individuals at four body sites: gut, tongue, right and left palm. After removing rows consisting of all 0's, the total number of different OTUs is 3131 for Person 2's gut data. Each feature is normalized to sum to one for all scenarios. 

For the rank 4 simulations, the fourth feature is chosen in the space spanned by the other three features and a unit vector with all elements the same, such that the perpendicular from the fourth feature to the plane of the other three features passes through the centroid of these three features. We vary the distance of this fourth type from the plane of the other three to assess the sensitivity of the methods in cases where one type is almost a linear combination of the other three.

For rank 2 simulations, the second feature is on the line in the space spanned by the first feature and a unit vector with all elements the same and in the direction from the first feature to the unit vector.

For rank 2 and rank 4, the further the distance is, the more distinct these features are. The distances are set to be 0, 0.0005 0.0008, 0.001, 0.0015 and 0.002. When the distance is 0, the feature matrix degenerates to a lower rank matrix. Thus, the rank 4 feature matrix degenerates to a rank 3 feature matrix and the rank 2 feature matrix becomes rank 1.

For all Poisson NMF data simulations, columns of the weight matrix are generated independently from a uniform distribution from 0 to 1. We adjust each column of the weight matrix to have a fixed sum sampled from the column sums for the healthy group of the Qin data for true NMF rank 2, 4, 6, 8 and 10. The weight matrix is rescaled to have fixed sum sampled from the column sums for Person 2's gut data of the moving picture data for true NMF rank 30. For microbiome data, these column sums are referred to as sequencing depth. Multiplying the feature matrix and the weight matrix, we get the parameters for the Poisson distribution. Each entry of the data matrix simulated for the Poisson NMF data is generated from a Poisson distribution with the Poisson mean given by the corresponding entry of $TW$. The estimated results of Boot-test, Decon-boot-test and NNLM in 50 replicates of different scenarios are summarized in Table \ref{tab2}, Table \ref{tab4} and Table \ref{tab6}. To compare the computational costs of the Boot-test and Decon-boot-test, we recorded the average time required to calculate the estimated rank for 50 replicates in each scenario in Table \ref{ttab1}, Table \ref{ttab2} and Table \ref{ttab3}. Both methods are coded in R (64-bit 4.0.3) and run on 1 node of the Graham Compute cluster of Canada with 1 Intel Xeon at 2.1Ghz with 2G of RAM and 1 core. The NNLM method is not compared here as it uses a different NMF algorithm while estimating the rank. The efficiency of the NMF algorithm is closely related to the computation cost of the rank selection method.

To compare the ability of Decon-boot-test and NNLM to detect Poisson data without NMF structure, we design a no NMF structure Poisson data simulation. The dimension of the no NMF structure Poisson data is the same as Person 1's gut data from the moving picture dataset, which has 1864 variables and 131 observations. In each replicate, each row of the mean of the no NMF structure Poisson data is generated from a log-multivariate Normal distribution. The mean of the multivariate Normal is a vector of length 1864 whose entries are independently simulated following a Normal distribution with mean 4 and standard deviation 3. For each sample from the multivariate Normal distribution, the mean is rescaled so that the sum of all means is equal to the log of the total sequencing depth from the corresponding sample in Person 1's gut data. The covariance matrix of the multivariate Normal is generated by 1864 eigenvalues evenly spaced from $3\times10^{-7}$ to 1 and orthogonal eigenvectors uniformly distributed over the 1863-dimensional sphere. If the Poisson data doesn't have NMF structure (which can be thought of as having rank equal to sample size), then rank selection methods should select a high rank (which doesn't satisfy the NMF rank selection rule $k<\frac{n\times p}{n+p}$ ~\cite{lee1999learning}). The average estimated ranks and standard deviations for the 50 replicates of the two methods are shown in Table \ref{tab6}.

 The Boot-test results in these tables indicate that for Poisson NMF data, the accuracy of Boot-test estimates increases with the number of initial values used for each NMF when $d=0.0005$ and $d=0.0008$ and is stable in most other cases. The computation time of Boot-test is positively related to the number of different initial values used for each NMF ($m$).  Decon-boot-test's computation time is between the time spent by $m=3$ and $m=10$ Boot-test. But Decon-boot-test estimates the ranks more accurately than Boot-test in all scenarios except for true rank 4 with $d=0.0005$. In Table \ref{tab2} and Table \ref{tab4}, the Decon-boot-test selects the true ranks in more replicates than NNLM especially when the distance from one feature to other features is small, for example $d=0.0005$ and $d=0.0008$. In Table \ref{tab6}, Decon-boot-test is comparable with the NNLM method when the NMF rank is low. For the high rank case, summaries of the estimated ranks show that both methods underestimate the rank, but Decon-boot-test gives a fairly close estimate for the rank, while NNLM estimates a very low rank. When there is no NMF structure, Decon-boot-test selects a high rank, while NNLM selects 1 for all replicates, which is somewhat misleading.


\begin{table}[htbp]%
\centering
\caption{Total number of times the true rank is selected out of 50 replicates and the 50 rank estimates' averages and standard deviations when the true rank is 2 for Poisson NMF data when bootstrap size is 50.\label{tab2}}%
\resizebox{\textwidth}{!}{
\begin{tabular}{ccabababababab}
\toprule
\rowcolor{LightCyan}
\textbf{d}     & & \multicolumn{2}{c}{0} & \multicolumn{2}{c}{0.0005} &  \multicolumn{2}{c}{0.0008} &  \multicolumn{2}{c}{0.001}   & \multicolumn{2}{c}{0.0015} &  \multicolumn{2}{c}{0.002}    \\

\midrule
    &$m=1$ & 47 & 1.06(0.24)  & 17 & 1.54(0.68)& 22 &2.04(0.75) &31 & 2(0.62)& 44 & 2(0.35)& 46&2.06(0.37)\\
    &$m=3$ & 46  &1.08(0.27) & 28 &2.16(0.65) & 39 &2.1(0.46) & 45 &2.12(0.39) & 40 & 2.2(0.40)& 42&2.16(0.37)\\
  \textbf{Boot-test}&$m=10$ & 47 &1.06(0.24) & 43 &2.2(0.53) & 45&2.08(0.28) & 44&2.06(0.25) & 43 &2.14(0.41)& 47 &2.06(0.24)\\
    &$m=30$ & 47 &1.06(0.23)  & 43 &2.16(0.42) & 45 &2.12(0.39)& 44&2.12(0.33) & 46&2.08(0.27) & 45&2.1(0.30)\\
    &$m=50$ & 47 &1.06(0.24)& 40&2.28(0.54) & 46&2.08(0.27)&44&2.14(0.40) & 45&2.16(0.55) & 45&2.04(0.20)\\
    \textbf{Decon-boot-test} & & \cellcolor{yellow}{50}&1(0) & \cellcolor{yellow}{50} &2(0)& \cellcolor{yellow}{50}&2(0)& \cellcolor{yellow}{50}&2(0) & \cellcolor{yellow}{50} &2(0) & \cellcolor{yellow}{50}&2(0) \\
    \textbf{NNLM} & & \cellcolor{yellow}{50}&1(0) & 39&2.12(0.52) & 43&2.02(0.38) & 44&2.04(0.35) & 48&2.04(0.18) & 47&2.06(0.24) \\
\bottomrule
\end{tabular}
}
\begin{tablenotes}
\item m: number of different initial values used for each NMF.
\item d: distance from one feature to a linear combination of other features.
\item First column for each value of d in grey: the number of times the true rank is selected out of 50 replicates. 
\item Second column for each value of d: the average of the 50 estimated ranks (before the bracket) and the standard deviation of the 50 estimates (in the bracket).
\item The most accurate estimates are highlighted in yellow.
\end{tablenotes}
\end{table}

\begin{center}
\begin{table}[t]%
\centering
\caption{Average computation time of the 50 replicates in hours when the true rank is 2 for Poisson NMF data when bootstrap size is 50.\label{ttab1}}%
\begin{tabular}{cccccccc}
\rowcolor{LightCyan}
\toprule
\textbf{d}  &   & 0   & 0.0005 & 0.0008 & 0.001 & 0.0015 & 0.002 \\
\midrule
    &$m=1$ & 0.2  & 0.3   & 0.4 & 0.5 & 0.4  &  0.5  \\
    &$m=3$ & 0.8  & 1.2  & 1 & 1.3 & 1.7  &  1.6  \\
    \textbf{Boot-test}&$m=10$ & 1.8 & 4.6 & 3.3 & 3.1 & 4.2  &  3.3  \\
    &$m=30$ & 5.3 & 9.3  &9.5 &9 &  8.5 & 9.2  \\
    &$m=50$ & 9  &  22.8  &17.5 & 18.4& 19  & 17.2  \\
    \textbf{Decon-boot-test}& & 1    & 2 & 1.3 & 1.4  & 2 & 2 \\
\bottomrule
\end{tabular}
\begin{tablenotes}
\item m: number of different initial values used for each NMF.
\item d: distance from one feature to a linear combination of other features.
\end{tablenotes}
\end{table}
\end{center}

\begin{center}
\begin{table}[htbp]%
\centering
\caption{Total number of times the true rank is selected out of 50 replicates and the 50 rank estimates' averages and standard deviations when the true rank is 4 for Poisson NMF data when bootstrap size is 50.\label{tab4}}%
\resizebox{\textwidth}{!}{
\begin{tabular}{ccabababababab}
\toprule
\rowcolor{LightCyan}
\textbf{d}     & & \multicolumn{2}{c}{0} & \multicolumn{2}{c}{0.0005} &  \multicolumn{2}{c}{0.0008} &  \multicolumn{2}{c}{0.001}   & \multicolumn{2}{c}{0.0015} &  \multicolumn{2}{c}{0.002}\\

\midrule
    &$m=1$ & 45 & 3.1(0.30) & 10&3.28(0.54)  & 14&3.66(0.85) & 18&4(0.81) &  35&4.2(0.57) & 42&4.14(0.45)\\
    &$m=3$ & 45 &3.12(0.39)  & 16 &3.48(0.65) & 31&4.24(0.62) & 38&4.22(0.51) & 43&4.14(0.35) & 43&4.14(0.35)\\
   \textbf{Boot-test}& $m=10$ & 39 &3.22(0.42) & 30 &3.92(0.63) & 34&4.26(0.44) & 34&4.28(0.50) & 39&4.17(0.38) & 42 &4.16(0.37)\\
    &$m=30$ & 40 &3.24(0.52) & 35&4.24(0.69)  &39 &4.28(0.57)& 41&4.2(0.45) &  42&4.16(0.37)& 41&4.18(0.39)\\
    &$m=50$ & 42 &3.22(0.58)& \cellcolor{yellow}{39} &4.24(0.59)&38&4.28(0.54) &37&4.28(0.50) & 42&4.2(0.49) & 43&4.16(0.42)\\
    \textbf{Decon-boot-test}& & \cellcolor{yellow}{50}&3(0) & 27& 3.54(0.50)& \cellcolor{yellow}{50}&4(0)& \cellcolor{yellow}{50}&4(0) & \cellcolor{yellow}{50} &4(0) & \cellcolor{yellow}{50}&4(0) \\
    \textbf{NNLM}& & \cellcolor{yellow}{50} &3(0)& 18&3.76(0.77) & 37&4.14(0.50) & 43&4.06(0.37) & 49 &4.02(0.14)& 48&4.04(0.20) \\
\bottomrule
\end{tabular}
}
\begin{tablenotes}
\item m: number of different initial values used for each NMF.
\item d: distance from one feature to a linear combination of other features.
\item First column for each value of d in grey: the number of times the true rank is selected out of 50 replicates. 
\item Second column for each value of d: the average of the 50 estimated ranks (before the bracket) and the standard deviation of the 50 estimates (in the bracket).
\item The most accurate estimates are highlighted in yellow.
\end{tablenotes}
\end{table}
\end{center}

\begin{center}
\begin{table}[t]%
\centering
\caption{Average computation time of the 50 replicates in hours when the true rank is 4 for Poisson NMF data when bootstrap size is 50.\label{ttab2}}%
\begin{tabular}{cccccccc}
\rowcolor{LightCyan}
\toprule
\textbf{d}  &   & 0   & 0.0005 & 0.0008 & 0.001 & 0.0015 & 0.002 \\
\midrule
    &$m=1$ & 0.6  & 0.7 & 0.8 & 0.8 & 1 &  1.2  \\
    &$m=3$ & 2  & 2.1   & 2.6 & 2.5 & 2.2  &  2.2  \\
    \textbf{Boot-test}&$m=10$ & 9  & 9.4  & 13.2 & 13.4 & 12  &  11.8  \\
    &$m=30$ &26.4  & 31.2  &32.4 & 28.4 & 27.7 & 28 \\
    &$m=50$  & 44 & 53 &55.3 & 55.1& 56.2  &  54\\
    \textbf{Decon-boot-test}& & 6.4 & 8.6 & 10.5 & 9&  11.2 & 10.4 \\
\bottomrule
\end{tabular}
\begin{tablenotes}
\item m: number of different initial values used for each NMF.
\item d: distance from one feature to a linear combination of other features.
\end{tablenotes}
\end{table}
\end{center}

\begin{center}
\begin{table}[htbp]%
\centering
\caption{Total number of times the true rank is selected of 50 replicates and the 50 rank estimates' averages and standard deviations when the true rank is 6, 8 and 10 for Poisson NMF data when bootstrap size is 50.\label{tab6}}%
\begin{tabular}{ccababababab}
\toprule
\rowcolor{LightCyan}
\multicolumn{2}{c@{}}{\textbf{rank}}   & \multicolumn{2}{c}{6}  & \multicolumn{2}{c}{8} &\multicolumn{2}{c}{10} & \multicolumn{2}{c}{30}&\multicolumn{2}{c}{nonNMF}\\

\midrule
&$m=1$ &   43  &6.16(0.42) & 41  & 8.18(0.39)& 36 &10.24(0.48) & & & & \\
&$m=3$&  43&6.14(0.35)  & 38 &8.24(0.43) &41&10.18(0.39)& & & & \\
\textbf{Boot-test}&$m=10$ & 41&6.18(0.39) & 39 &8.22(0.42)& 44 &10.12(0.33)& & & & \\
& $m=30$ & 34 &6.32(0.47)&  37 &8.26(0.44) & 38&10.24(0.43)& & & & \\
&$m=50$ &  38 &3.32(2.95) &  34 &8.32(0.47) & 40&10.2(0.40)& & & & \\
 \textbf{Decon-boot-test} & & \cellcolor{yellow}{50}  &6(0) &   \cellcolor{yellow}{50} &8(0)  & 47&10.06(0.24)& 0 &\cellcolor{yellow}{25.04(1.4)}&  &\cellcolor{yellow}{39.69(8.61)}\\
\textbf{NNLM} & & \cellcolor{yellow}{50} &6(0)& 45&8.1(0.30) & \cellcolor{yellow}{50} &10(0)& 0 &2.76(1.76)& &1(0)\\
\bottomrule
\end{tabular}
\begin{tablenotes}
\item m: number of different initial values used for each NMF.
\item First column under each rank in grey: the number of times the true rank is selected out of 50 replicates. (first column is not available for nonNMF data) 
\item Second column under each rank: the average of the 50 estimated ranks (before the bracket) and the standard deviation of the 50 estimates (in the bracket).
\item The most accurate estimates are highlighted in yellow.\end{tablenotes}
\end{table}
\end{center}

\begin{center}
\begin{table}[t]%
\centering
\caption{Average computation time of the 50 replicates in hours when the true rank is 6, 8 and 10 for Poisson NMF data when bootstrap size is 50.\label{ttab3}}%
\begin{tabular}{ccccc}
\rowcolor{LightCyan}
\toprule
\textbf{rank}  &   & 6   & 8 & 10\\
\midrule
    &$m=1$ &  2.8 & 3.6 & 6  \\
    &$m=3$ & 6  &  11.5  & 20 \\
    \textbf{Boot-test}&$m=10$ &  19.2 & 37  & 55.2\\
    &$m=30$ & 63.9 & 102  & 207.6\\
    &$m=50$  & 101.3 & 209.9 & 323.5\\
    \textbf{Decon-boot-test}& & 8.5 & 33 & 35.5\\
\bottomrule
\end{tabular}
\begin{tablenotes}
\item m: number of different initial values used for each NMF.
\end{tablenotes}
\end{table}
\end{center}


\subsection{Normal data simulation}\label{subsec32}

We generate Normal NMF data with true rank equal to 2, 3, 4, 6, 8 and 10. For each scenario, we simulate 50 replicates. 

Each element in the feature matrix for the Normal NMF data is independently sampled from a gamma distribution with shape parameter 3 and rate parameter 2, then multiplied by a Bernoulli random variable with parameter $0.7$ to control the sparsity of the feature matrix. All replicates in the same scenario share the same feature matrix. The dimension of the features is 2780 initially. After removing rows with all zero elements in the feature matrix, there are 2544 rows remaining in the type 2 feature matrix and 2721 rows in the feature matrices for other ranks.

The 100 columns of the weight matrix $W$ are generated independently from a Dirichlet distribution with parameters all set to be $1.5$ and rescaled by a factor of 10.

Each entry of the data matrix is simulated following a Normal distribution with mean given by the corresponding entry in the matrix product $TW$ and variance 1. Negative entries in the data are replaced by 0 to generate nonnegative data. 

Table \ref{tab7} shows for Normal NMF data, the boot-test's accuracy increases when more sets of initial values are used for each NMF. Both the Decon-boot-test and NNLM estimate the true ranks for Normal NMF data more accurately than Boot-test. Decon-boot-test and NNLM are both extremely accurate at estimating the rank for all scenarios. Table \ref{ttab4} shows the average time required by Decon-boot-test to calculate the estimated rank for the 50 replicates in each Normal NMF data scenario is usually between the computational expense of $m=3$ and $m=10$ Boot-test. 

\begin{center}
\begin{table}[htbp]%
\centering
\caption{Total number of times the true rank is selected out of 50 replicates for Normal NMF data and the 50 rank estimates' averages and standard deviations when bootstrap size is 50.\label{tab7}}%
\resizebox{\textwidth}{!}{
\begin{tabular}{ccabababababab}
\toprule
\rowcolor{LightCyan}
    \multicolumn{2}{c}{\textbf{rank}} &  \multicolumn{2}{c}{2} & \multicolumn{2}{c}{3} & \multicolumn{2}{c}{4} & \multicolumn{2}{c}{6} & \multicolumn{2}{c}{8} & \multicolumn{2}{c}{10} \\
\midrule
    &$m=1$ & 47 &2.1(0.46) & 40& 3.02(0.14)& 39& 4.22(0.42)& 40& 6.2(0.40)& 42& 8.16(0.37)& 43&10.14(0.35)\\
    &$m=3$& 41 &2.22(0.55)& 38&3.24(0.43) &42&4.18(0.44) & 44&6.12(0.33) & 41&8.18(0.39) & 46&10.08(0.27) \\
    \textbf{Boot-test}&$m=10$ & 42&2.2(0.49) & 40&3.12(0.33) & 42&4.16(0.37) & 42&6.16(0.37) & 47&8.06(0.24) & 41&10.18(0.39) \\
    &$m=30$ & 44 &2.14(0.40)& 40&3.22(0.46) & 42&4.16(0.37) & 43&6.16(0.42)) & 45&8.1(0.30) & 43&10.14(0.35) \\
    &$m=50$ & 46&2.08(0.27) & 42&3.16(0.37) & 46&4.08(0.27) & 43 &6.14(0.35)& 42&8.16(0.37) & 42&10.16(0.37)\\
    \textbf{Decon-boot-test} && \cellcolor{yellow}{50}&2(0) & 49 &3.02(0.14)& \cellcolor{yellow}{50}&4(0) & \cellcolor{yellow}{50} & 6(0)&49&8.02(0.14) & \cellcolor{yellow}{50}&10(0)\\
    \textbf{NNLM} && \cellcolor{yellow}{50}&2(0) & \cellcolor{yellow}{50} &3(0)& \cellcolor{yellow}{50}&4(0) & \cellcolor{yellow}{50} &6(0)& \cellcolor{yellow}{50}&8(0) & \cellcolor{yellow}{50}&10(0)\\
\bottomrule
\end{tabular}
}
\begin{tablenotes}
\item m: number of different initial values used for each NMF.
\item First column under each rank in grey: the number of times the true rank is selected out of 50 replicates. 
\item Second column under each rank: the average of the 50 estimated ranks (before the bracket) and the standard deviation of the 50 estimates (in the bracket).
\item The most accurate estimates are labeled as yellow.
\end{tablenotes}
\end{table}
\end{center}

\begin{center}
\begin{table}[t]%
\centering
\caption{Average computation time of the 50 replicates in hours for Normal NMF data when bootstrap size is 50.\label{ttab4}}%
\begin{tabular}{cccccccc}
\rowcolor{LightCyan}
\toprule
\textbf{rank}  &   & 2   & 3 & 4 & 6 & 8 & 10 \\
\midrule
    &$m=1$ & 0.1  & 0.5 & 0.5 & 1 & 2.1 &  3.5  \\
    &$m=3$ & 0.4  &  1.3  & 1.3 & 3 & 7.8  &  10.2  \\
    \textbf{Boot-test}&$m=10$ & 2.7  & 2.5  & 4.2 & 10 & 25.8  &  33.5  \\
    &$m=30$ & 3.6 & 6.3  &11.7 &28.7 & 69.7  & 95.7 \\
    &$m=50$  & 5.7 & 12.9 & 23& 53.6&  104 &  186.5\\
    \textbf{Decon-boot-test}& & 2 & 2.5 & 3 & 8.3&  23.5 & 33.5 \\
\bottomrule
\end{tabular}
\begin{tablenotes}
\item m: number of different initial values used for each NMF.
\end{tablenotes}
\end{table}
\end{center}

\subsection{Conclusion of simulations}\label{subsec34}

The simulation results show that Decon-boot-test is better than Boot-test in estimation accuracy for both Poisson data and Normal data. By using a large number of starting points, Boot-test can achieve similar accuracy to Decon-boot-test, but is far more computationally expensive with no benefit in accuracy. When the true rank is small, the performance of Decon-boot-test for Normal data is comparable with NNML, and for Poisson data, its performance is better than NNML, especially when the features are hard to distinguish from each other. When the rank is large, both methods fail to find the true rank but Decon-boot-test's estimated rank is much closer to the true rank. When the data is not NMF structured, Decon-boot-test selects very large rank, while NNLM selects rank 1 for all replicates. Thus, Decon-boot-test gives a much clearer indication of the lack of NMF structure, which can be used to better interpret the data.

\section{Real Data application}\label{sec4}

We apply our NMF rank selection method to moving picture data \cite{Caporaso}. As mentioned in Section \ref{subsec31}, this dataset records the microbiome of two people's gut, tongue, left and right palm for a long time period. Person 1 was female, and was sampled over a period of 186 days, with a total of 131--135 samples for each body site. Person 2 was male and was sampled over a period of 446 days, with a total of 336--373 samples for each body site. The total number of
variables (different OTUs) across all samples was more than 15000.
In spite of this extensive sampling, no temporal core microbiome was detected, with
only a small subset of OTUs reoccurring at the same body site across
all time points \cite{Caporaso}. These data sets have previously been analysed using NMF \cite{cai2017learning}, but without a good way to select the rank, it is possible that there are undetected biologically important patterns in the data that can be discovered by reanalysing the data using the number of types selected by the Decon-boot-test method. We therefore apply our method to the gut data from both individuals to choose the number of types, and examine the fitted types and weights to obtain biological insights that were not observed in previous analyses of the data using a different number of types. Our method selects NMF rank as 12 for Person 1's gut data and NMF rank as 6 for Person 2's gut data. 

\begin{figure}[htbp]
\centering
\includegraphics[width=19cm,height=11cm]{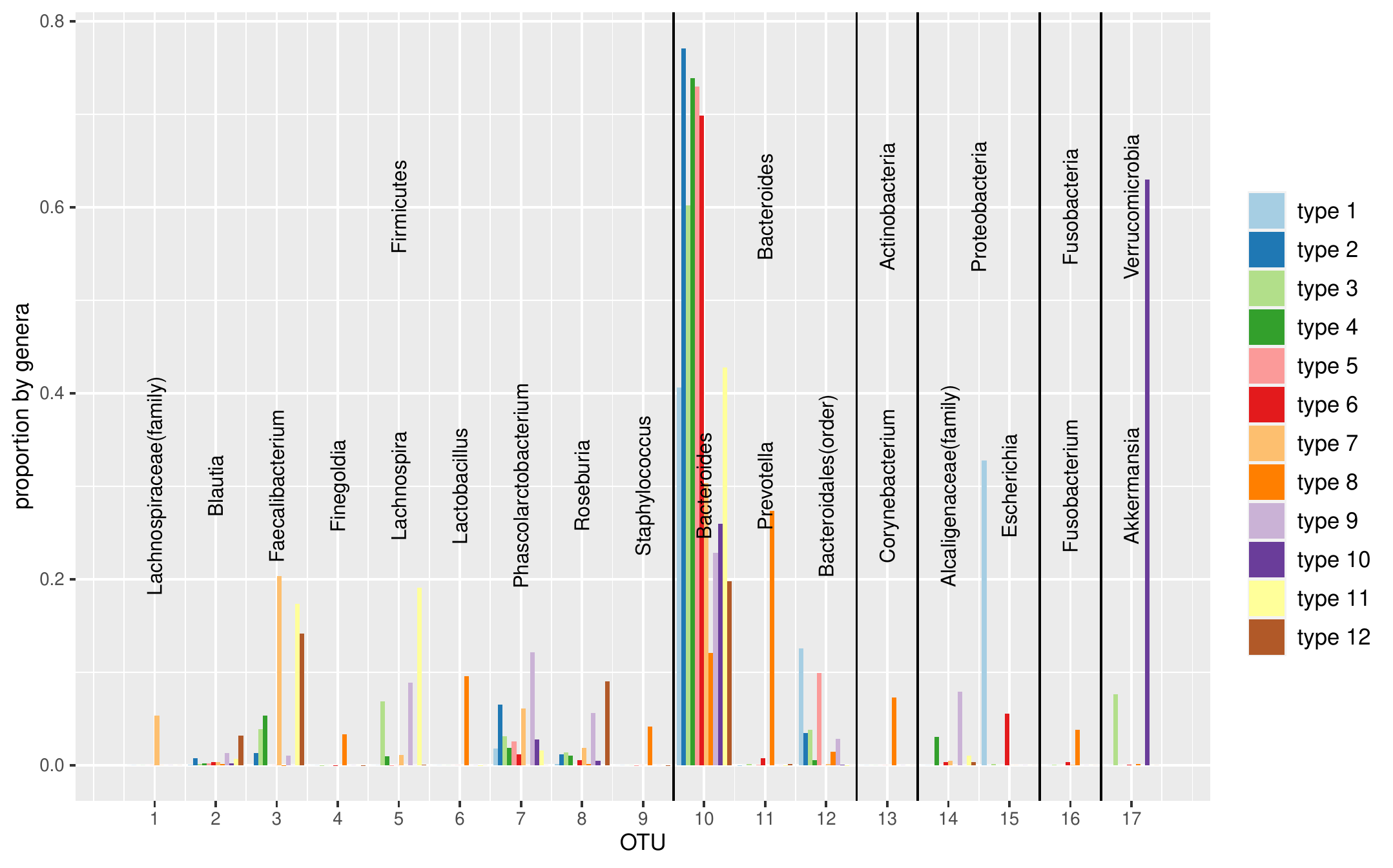}
\caption{Major genera for Person 1's gut feature matrix. The genera from the same phylum are in the same block which is labeled by their phylum names and the bars are labeled by the genus names or higher level of taxonomic rank if it's unclassified at genus level. Each bar is colored according to its type.}
\label{F1}
\end{figure}

\begin{figure}[htbp]
\centering
\subfigure{\includegraphics[width=18cm,height=10cm]{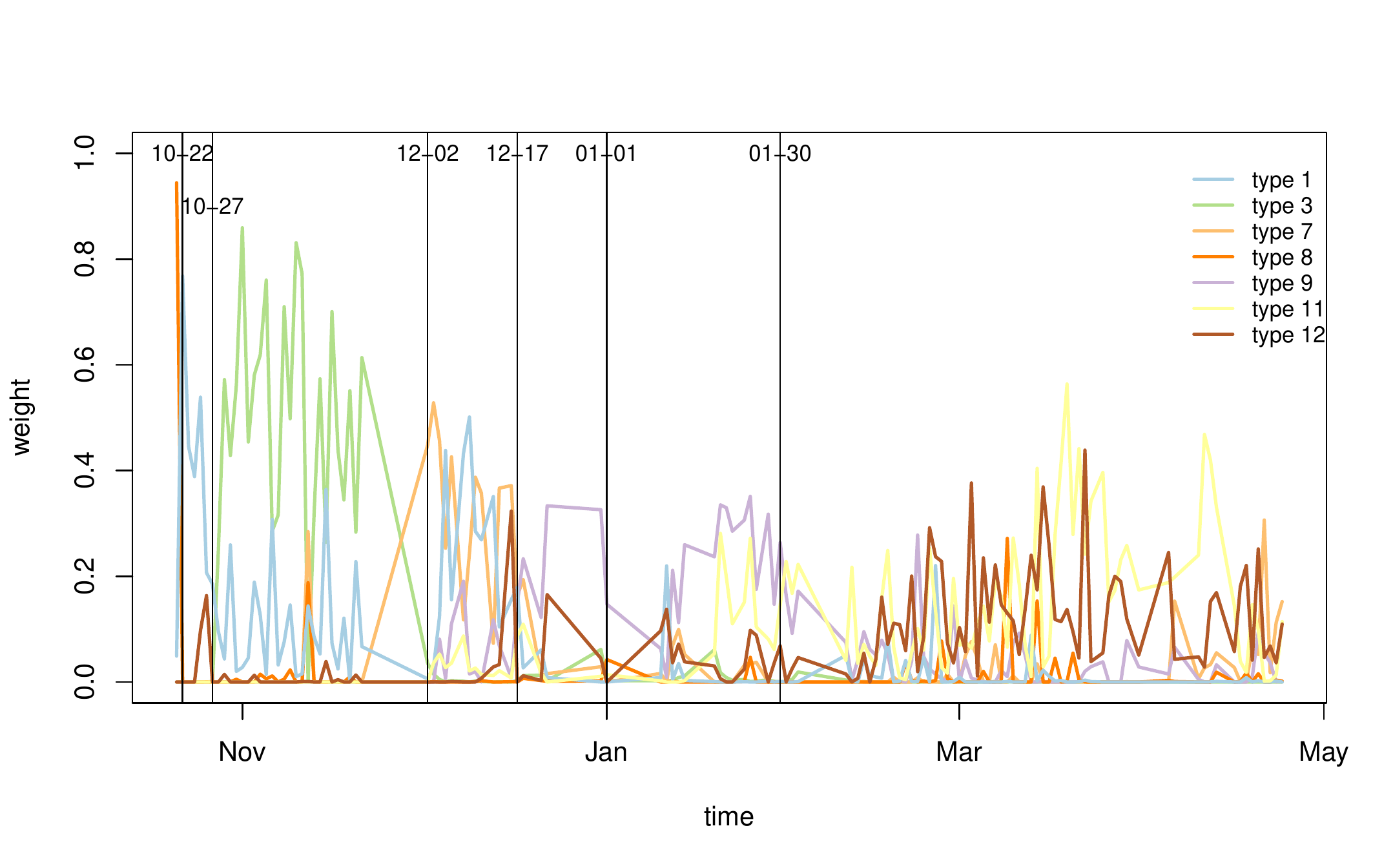}}
\subfigure{\includegraphics[width=18cm,height=10cm]{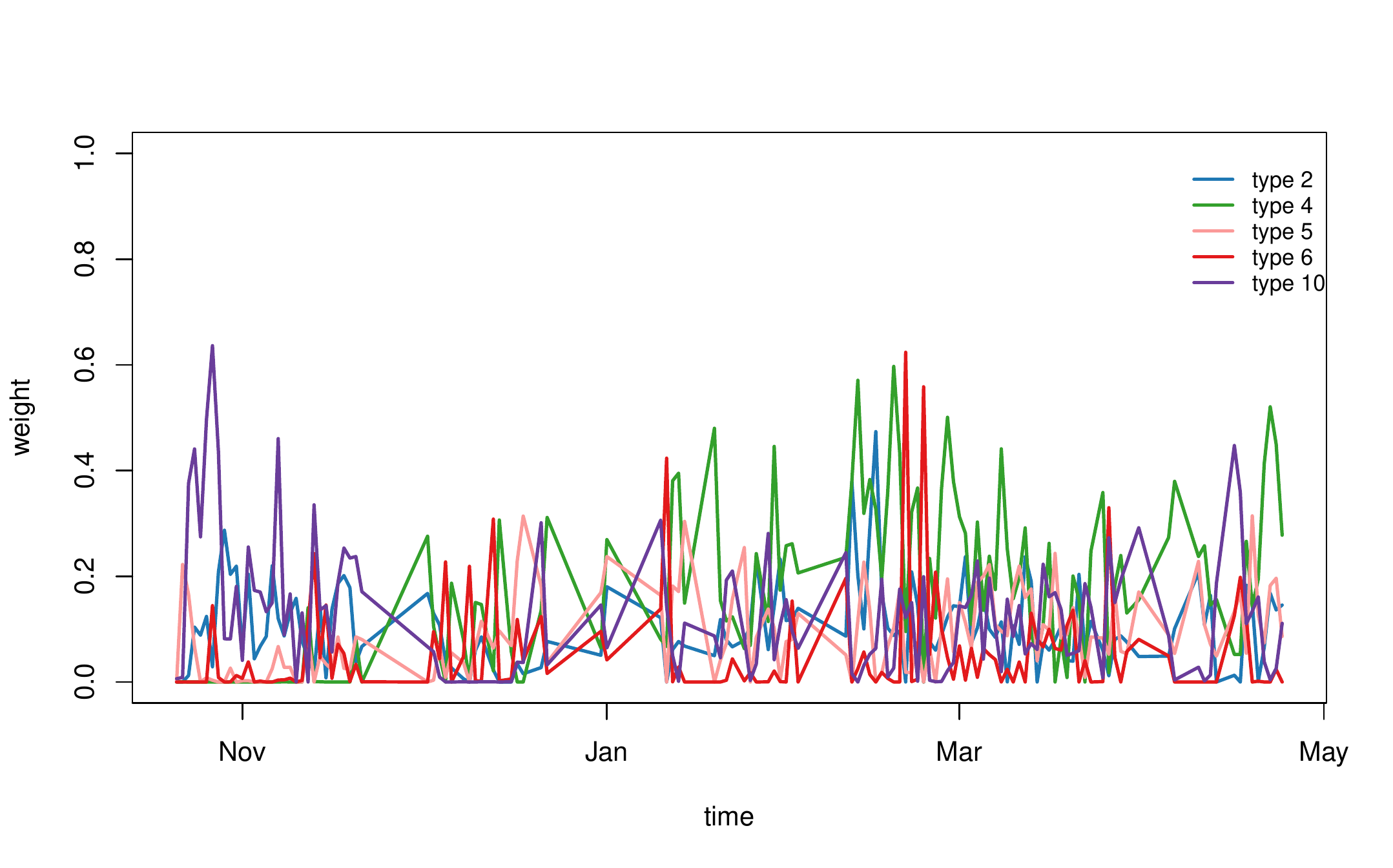}}
\caption{Gut weight matrix time series plot for Person 1. The
top plot shows Person 1's gut weight matrix on type 1, type 3, type 7, type 8, type 9, type 11 and type 12 from 12 rank NMF. The bottom plot shows the weight matrix for other types. Each weight is colored the same as its corresponding type.}
\label{F2}
\end{figure}

\begin{figure}[htbp]
\centering
\includegraphics[width=19cm,height=11cm]{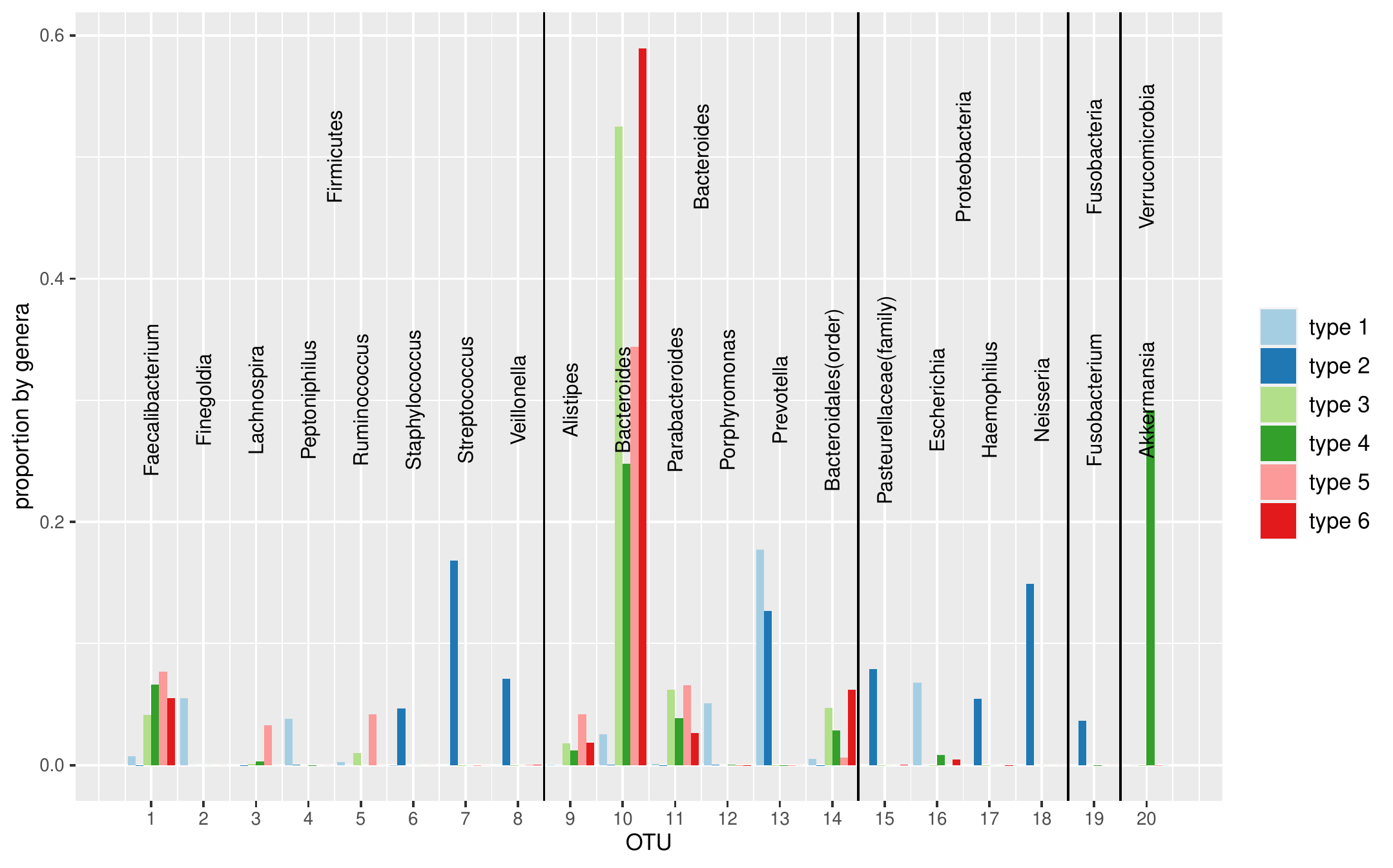}
\caption{Major genera for Person 2's gut feature matrix. The genera from the same phylum are in the same block which is labeled by their phylum name and the bars are labeled by the genus names or higher level of taxonomic rank if it's unclassified at genus level. Each bar is colored according to its type.}
\label{F3}
\end{figure}

\begin{figure}[htbp]
\centering
\subfigure{\includegraphics[width=18cm,height=10cm]{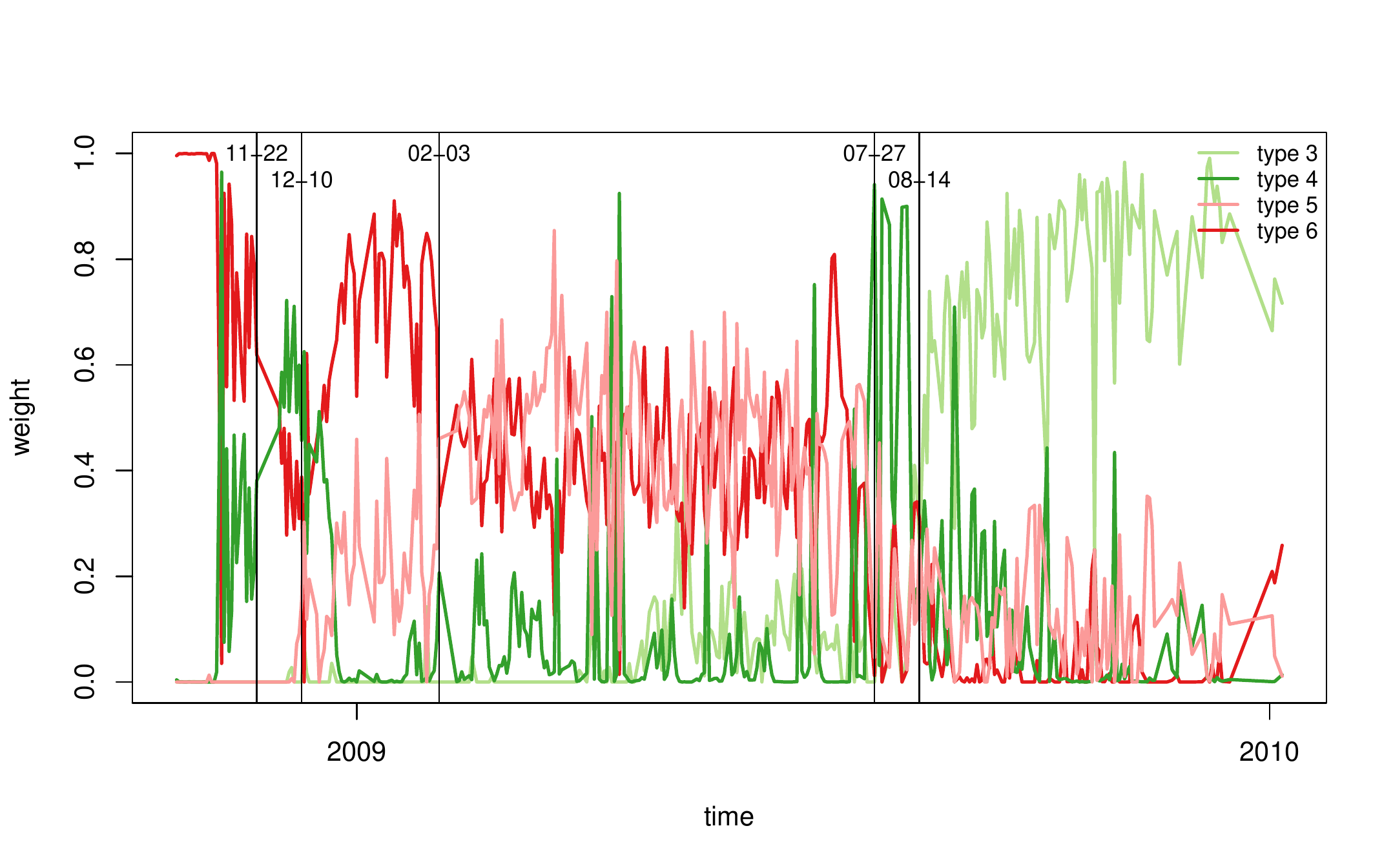}}
\subfigure{\includegraphics[width=18cm,height=10cm]{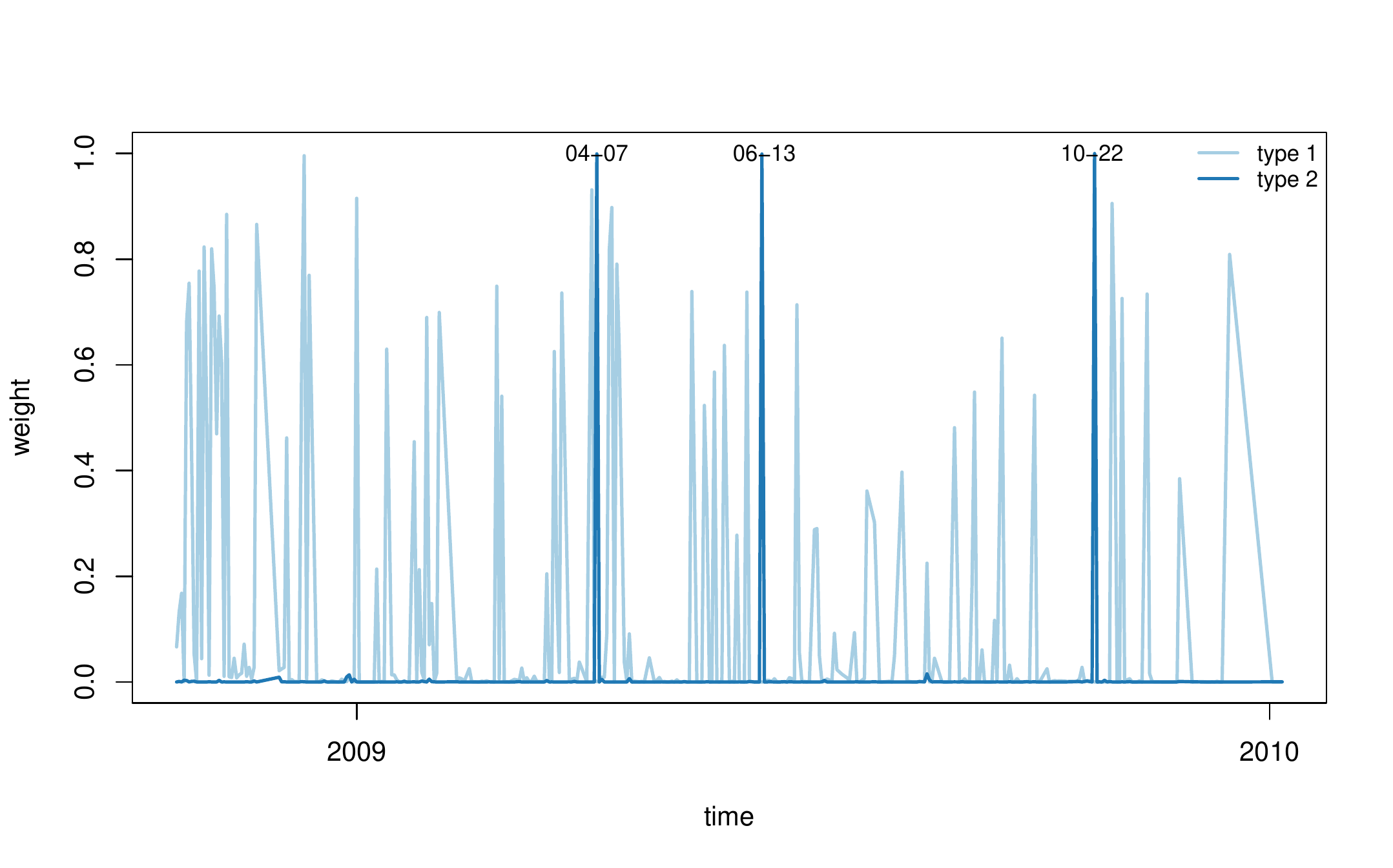}}
\caption{Gut weight matrix time series plot for Person 2. The
top plot shows Person 2's gut weight matrix normalized on type 3, type 4, type 5 and type 6 from 6 rank NMF. The bottom plot shows the weight matrix for other types. Each weight is colored the same as its corresponding type.}
\label{F4}
\end{figure}

To examine the microbiome community structure of the gut from the NMF results, we plot the relative abundance
of each genus in each feature in Figure \ref{F1} for Person 1's gut and Figure \ref{F3} for Person 2's gut. The elements of each feature sum to 1 so the coefficients can be interpreted as the proportions of each OTU in that feature. As the NMF results are usually highly sparse, we use a cut-off of $3\%$ for each type. That is, only those genera with OTUs above $3\%$ composition in at least one type are shown in the plot. The outstanding OTUs for features of both persons' guts are from the phyla Firmicutes, Bacteroides, Actinobacteria, Proteobacteria, Fusobacteria and Verrucomicrobia. Most of the features have abundant Firmicutes and Bacteroides which is consistent with the fact that Firmicutes and Bacteroidetes constitute $90\%$ of the gut microbiota, while Actinobacteria, Proteobacteria, Fusobacteria, and Verrucomicrobia contribute to the microbial population to a lesser extent\cite{scarpellini2015human}.

Figure \ref{F1} shows the abundance of the main genera in Person 1's gut features. We find only 30 out of more than eighteen hundred OTUs are larger than the cut-off of $3\%$. These 30 OTUs are from 17 genera. From the plot, we see Type 2, 4, 5 and 6 share similar compositions at genus level. They all have extremely high abundance of Bacteroides and much lower abundance of other genera. These types differ in the compositions of other genera. In particular, the secondary genus in the community is Faecalibacterium in Type 4, Phascolarctobacterium in Type 2, an unclassified genus from order Bacteroides in Type 5 and Escherichia in Type 6. Type 10 is dominated by Akkermansia rather than Bacteroides. As lean individuals have more Bacteroides, while obese individuals have more Firmicutes in their intestinal microbiota and the abundance of Akkermansia is negatively correlated to human's weight\cite{turnbaugh2006obesity}\cite{zhou2020gut}, Type 2, Type 4, Type 5, Type 6 and Type 10 may be healthy communities with regard to body weight. Type 1 mainly consists of Escherichia and an unclassified genus from order Bacteroidales. We know that a number of species of Escherichia are pathogenic\cite{hogan2010bacteria} so this type could be associated with infection with pathogens. Type 8's major genus is Prevotella whose increased abundance is linked to inflammatory disorders in emerging studies, suggesting that at least some strains exhibit pathobiontic properties\cite{larsen2017immune}. After an intervention with a Mediterranean diet and a low-fat diet, increased Prevotella and Bacteroides levels have also been observed\cite{precup2019gut}, so Type 8 could be associated with a short-term change of diet. Less abundant genera in Type 8 such as Finegoldia, Staphylococcus, Lactobacillus, Corynebacterium and Fusobacterium have, on rare occassions, been found to cause human infection\cite{cobo2021infections}\cite{lee2021staphylococcus}\cite{rossi2019members}\cite{bernard2012genus}\cite{afra2013incidence}. So Type 8 could also be associated with infection. Type 3 consists of Bacteroides, Lachnospira and Akkermansia. Type 7 has high proportions of Faecalibacterium and an unclassified genus from family Lachnospiraceae. Type 11 and Type 12 also have high proportions of Faecalibacterium while Type 11 has fairly abundant Lachnospira and Type 12 has fairly abundant Roseburia. Previous studies suggest that Roseburia and Lachnospira are strongly associated with vegetable diets, and also negatively associated with the omnivore diet\cite{vacca2020controversial}. Also, Faecalibacterium and Roseburia may be two major proponents of weight loss\cite{chakraborti2015new}. This suggests that Type 3, Type 7, Type 11 and Type 12 may all be related to low-fat diets. Type 9 has fairly abundant Phascolarctobacterium which has been found to increase under high fat diet and to be positively correlated to a positive mood and body weight\cite{li2016gut}\cite{lecomte2015changes}.

To investigate the temporal dynamics of the Person 1's gut microbiome, the
weight matrix is plotted in Figure \ref{F2}. The weights for each time point are normalized to sum to 1. We see that there is substantial short-term fluctuation for all types. Types 6 and 8 have consistently low weights, with the exception of single-day spikes. This would be consistent with Types 6 and 8 being associated with short-term infections. It could also be consistent with Type 8 being associated with diet, if the association is short-term, then the spikes might correspond to the individual eating particular foods. Type 9 is dominant with a relatively stable weight for around one week at end of December which would be consistent with the diet change and happy mood. Other types show elevated levels for time periods of weeks or months, indicating changes to the microbial community over time. This seems plausible dynamics for healthy types. 

Figure \ref{F3} shows the abundance of the main genera in Person 2's gut features. 33 out of 3131 OTUs have larger than $3\%$ proportions in at least one feature. These 33 OTUs are from 20 genera. Person 2's Type 3, Type 5 and Type 6 are similar. They are all dominated by Bacteroides with the next most abundant genera, Faecalibacterium and Parabacteroides, appearing in much smaller proportions. Type 4 has high proportions of Bacteroides and Akkermansia. Type 1 is more diverse, with no genus exceeding $20\%$ of the total abundance of this type. The most abundant genera are Prevotella, Escerichia, Finegolida and Porphyromonas. Many species from these genera have previously been associated with periodontitis\cite{mei2020porphyromonas}. Although the species in this dataset are not the ones previously associated with periodontitis, it seems plausible that this type might be associated with periodontic infections. Type 2 consists of Streptococcus (gordonii species and oralis species), Prevotella (nanceiensis species, melaninogenica species and an unclassified species), Neisseria (subflava species) and a lower abundance of Staphylococcus (epidermidis species), Veillonella (parvula species and an unclassified species), an unclassified genus from family Pasteurellaoeae, Haemophilus (parainfluenzae species) and Fusobacterium (unclassified species). The Prevotella species in Type 2 are different from the species in Type 1. S. gordonii, V. parvula and Fusobacterium are known to be opportunistic pathogens that can cause periodontitis\cite{park2020streptococcus}\cite{bongaerts2004isolation}\cite{broadley2017get}. Also, S. gordonii, V. parvula, H. parainfluenzae and N. subflava are recognized in opportunistic infections such as endocarditis and meningitis\cite{bongaerts2004isolation}\cite{broadley2017get}\cite{latyshev2013purulent}\cite{deza2016isolation}\cite{kaplan2002biofilm}. P. melaninogenica is also an important human pathogen in various anaerobic infections, often mixed with other aerobic and anaerobic bacteria\cite{brook2007anaerobic}. Thus, Type 2 might also be associated with infection.

Figure \ref{F4} shows the weights of each type in Person 2's gut over time. We see that Types 1 and 2 are usually very low abundance, but have occasional spikes. This is consistent with these types corresponding to infections. The other features show much more stability over time, having similar abundance levels over long periods of time (when normalised with respect to the total of these four types, with Types 1 and 2 excluded). There are some very abrupt changes to the abundance levels of these types at various times during the study. This is consistent with these types representing small variations on a healthy community.

\section{Conclusion}

We have developed an NMF rank selection method based on hypothesis testing using a deconvolved bootstrap to estimate the null distribution. The simulations show that our rank selection method can estimate the rank accurately for both Poisson and Normal NMF data when the true rank is small. When the NMF features are close or the true NMF rank is large, our method has better performance than NNLM. We applied our method to a microbiome data set. With the number of ranks selected, we were able to gain new insights beyond previous analyses of the same data. The new insights are biologically extremely plausible, and will hopefully lead to new research in the field. 

There are a number of directions for future research. Firstly, the sequential nature of the test means that we need to recompute the null distribution for each rank, which is computationally very expensive. The test could be made computationally more efficient for large ranks by using a more efficient search. Instead of increasing the tested rank by 1 each time, we could increase it by more. If the test does not reject the null hypothesis, it would then be necessary to perform the tests for lower ranks. This would increase computation when the rank is low, but could decrease it significantly for higher rank. More heuristics could be added to decide which hypothesis tests to perform for maximum efficiency.

Another issue is that a single failure to reject the null hypothesis can cause the method to select the current rank. It might be possible to develop a more robust method that combines the output of additional hypothesis tests to estimate the rank more reliably.

Another direction for future research is estimation of variance for the parametric bootstrap in the Normal case. The variance is estimated from the residuals of the model. However, because the model is fitted to the data, these residuals will be smaller when the rank is larger. For linear regression, there is a correction to get an unbiased estimate for the variance. However, for NMF, the nonnegativity constraint means that this correction is not applicable, so another method is needed to obtain a more stable estimate for the variance. This will make the critical value the test based on from the parametric bootstrap more accurate. 

The application of deconvolution to deal with optimization errors in bootstrap samples has potential applications in any field where full optimization is computationally expensive. This is common for discrete optimization problems such as variable selection, clustering, phylogenetics and many other areas. Applying the deconvolved bootstrap in these areas could prove a very fruitful topic for future research.

\subsection*{Author contributions}

All three authors contributed to the development of methodology, design of the simulations, interpretation of the real data analysis, and writing of the paper. All three authors contributed to the implementation of the method. All authors read and approved the final manuscript.

\subsection*{Financial disclosure}

The second author is supported by NSERC grant RGPIN-2017-05108. The third author is supported by NSERC grant RGPIN/04945-2014.

\subsection*{Conflict of interest}

The authors declare no potential conflict of interests.

\section*{DATA AVAILABILITY STATEMENT}
The data analysed in this manuscript have all been previously published in other papers.

The Moving picture data set is in Qiita study 550.

The Qin data set is available in the format used for our analysis with the BiomeNet package at  \url{http://sourceforge.net/projects/biomenet/}.

\bibliography{wileyNJD-AMA}

\clearpage

\end{document}